\documentclass[12pt]{article}
\usepackage{epsfig}

\usepackage{graphicx}
\usepackage{xcolor}
\usepackage{cancel}
\usepackage{amssymb,amsmath}

\def\harr#1#2{\smash{\mathop{\hbox to .3in{\rightarrowfill}}
 \limits^{\scriptstyle#1}_{\scriptstyle#2}}}

\def\s2{\frac{1}{\sqrt2}}

\def\beqa{\begin{eqnarray}}
\def\eeqa{\end{eqnarray}}

\def\Dsl{\,\raise.15ex\hbox{/}\mkern-13.5mu D} 
\def\d3{d^3}

\def\F{{\cal F}}

\def\a{\alpha'}

\newcommand{\be}{\begin{equation}}
\newcommand{\ee}{\end{equation}}
\newcommand{\ba}{\begin{eqnarray}}
\newcommand{\ea}{\end{eqnarray}}

\newcommand{\bs}{\be \begin{split}}
\newcommand{\ens}{ \end{split} \ee}
\newcommand{\nn}{\nonumber}

\newcommand{\qp}{\mathbb{Q}_p}

\newcommand{\st}{\mathrm{sgn}_\tau}
\newcommand{\pint}{\int_{\mathbb{Q}_p}}
\newcommand{\fer}{{\scriptscriptstyle F}}
\newcommand{\bos}{{\scriptscriptstyle B}}
\newcommand{\G}{\mathcal{G}}

\usepackage{amssymb,amsmath}

\usepackage{hyperref}
\hypersetup{
        colorlinks=true,
        linkcolor=blue,
        citecolor=red,
}

\def\be{\begin{equation}}
\def\ee{\end{equation}}
\def\beqa{\begin{eqnarray}}
\def\eeqa{\end{eqnarray}}

\begin{document}

\begin{center}
\Large{\bf Towards Non-Archimedean Superstrings} \vspace{0.5cm}

\large  Hugo Garc\'{\i}a-Compe\'an\footnote{e-mail address: {\tt
compean@fis.cinvestav.mx}}, Edgar Y. L\'opez\footnote{e-mail
address: {\tt elopez@fis.cinvestav.mx}}

\vspace{0.3cm}

{\small \em Departamento de F\'{\i}sica, Centro de
Investigaci\'on y de Estudios Avanzados del IPN}\\
{\small\em P.O. Box 14-740, CP. 07000, Ciudad de M\'exico, M\'exico}\\

\vspace*{1.5cm}
\end{center}

\begin{abstract}
An action for a prospect of a $p$-adic open superstring on a target Minkowski space is proposed.
The action is constructed for `worldsheet' fields taking values in
the $p$-adic field $\mathbb{Q}_p$, but it is assumed to be obtained from
a discrete action on the Bruhat-Tits tree. This action is proven to have an analogue of
worldsheet supersymmetry and the superspace action is also constructed in terms of superfields. 
The action does not have conformal symmetry, however it is implemented in the definition of the amplitudes. The tree-level amplitudes for this theory are obtained for $N$ vertex operators
corresponding to tachyon superfields and a Koba-Nielsen formula is
obtained. Finally, four-point amplitudes are computed explicitly and
they are compared to previous work on $p$-adic superstring
amplitudes.

\vskip 1truecm

\end{abstract}

\bigskip

\newpage

\section{Introduction}
\label{sec:intro}

String theory is a very strong candidate for a quantum theory of gravity. Its
non-perturbative formulation known as AdS/CFT correspondence has
been successfully applied to many systems of gravity and field
theory \cite{Aharony:1999ti}. Among its most remarkable applications
is that of describing the quantum properties of black holes
constructed from D-brane configurations. This feature is achieved by
counting the corresponding open-string states of the supersymmetric brane
configurations. There are other phenomena also described in terms of
brane configurations as the loss of information and the emergence of
spacetime itself from entangled states in the dual conformal field
theory. Some of these considerations are in an early stage and it is
observed that in the standard correspondence is quite difficult to find
some progress due to the complicated nature of the calculations. Thus, some simpler models that capture some essential
features of the AdS/CFT correspondence are very important to be
explored in order to achieve some progress. Recently a model was
proposed in \cite{Gubser:2016guj,Heydeman:2016ldy}, regarding a
$p$-adic version of AdS/CFT correspondence. These $p$-adic models
capture the essential features of the usual correspondence. The bulk
is described on the Bruhat-Tits tree and its boundary field theory
is given in terms of a $p$-adic CFT on the line proposed several years
ago \cite{Melzer:1988he}.

The study of a $p$-adic string theory was introduced many years ago
and further developed in
\cite{Volovich:1987nq,Freund:1987kt,Freund:1987ck,Brekke:1987ptq,Frampton:1987sp}.
In these references it was studied the proposal of considering the
tree-level amplitudes of a bosonic open string theory with the fields on
the worldsheet taking values over the $p$-adic number field
$\mathbb{Q}_p$. Some reviews on non-Archimedean string amplitudes
can be found in Refs.
\cite{Hlousek:1988vu,Brekke:1993gf,Vladimirov:1994wi,Bocardo-Gaspar:2021ukf}.

Later these amplitudes were derived from a $p$-adic worldsheet
action defined on the Bruhat-Tits tree \cite{Zabrodin} or projected
out as an effective theory to the boundary of the tree, the $p$-adic line $\mathbb{Q}_p$,
in terms of the Vladimirov derivative \cite{Spokoiny}. Both methods
were found to be equivalent in order to derive the mentioned open
string amplitudes. In this context the study of the regularization
of the amplitudes for non-Archimedean open strings in terms of local zeta functions was discussed  in 
\cite{Bocardo-Gaspar:2016zwx}. The rigorous study of open and closed strings on any local field of characteristic zero was 
considered in \cite{Bocardo-Gaspar:2019pzk}, and the limit $p\to 1$ was described in
\cite{Bocardo-Gaspar:2017atv}. The discussion on the incorporation of a
constant NS $B$-field is studied in Ref. \cite{GhoshalBfield}. The
regularization of the Ghoshal and Kawano amplitudes was carried out
in \cite{OurB-Field}. Very recently the procedure followed in
\cite{Spokoiny,Zabrodin} has been studied from a rigorous point of
view in \cite{Fuquen-Tibata:2021vnm}.

The study of tree-level
amplitudes in superstring theory was considered independently in
Refs. \cite{Arefeva:1988qi,MarshakovNew}. Direct analogues for the 4-point superstring amplitudes were considered in \cite{Ruelle:1989,Brekke:1993gf}. In order to add fermions in the $p$-adic string amplitudes we require of extending the non-Archimedean formalism to 
include Grassmann numbers. Some further developments
of the formalism were carried out in \cite{Dragovich:2005we}.

In the context of $p$-adic AdS/CFT some results concerning the
study of non-Archimedean versions of fermionic systems as SYK melonic
theories were obtained in \cite{GubserMelonic}. Furthermore a way
of introducing the spin was proposed in Ref. \cite{Gubser:2018cha}.
Motivated in part by these works, in \cite{QuGaoSpinor} was
studied the fermionic field theory on the Bruhat-Tits tree and its
effective action on the boundary.

In the present article, motivated by all the mentioned works, we
propose a worldsheet action containing bosons and fermions on the
$p$-adic worldsheet projected on the boundary. We will show that
this action is supersymmetric and thus might be considered as a
$p$-adic analogue of the worldsheet superstring action in the
superconformal gauge \cite{GSWSuperstrings}. Moreover we will show
that this action can be rewritten as an action in a $p$-adic version
of the ordinary superspace. Furthermore we compute the tree-level
$N$-point open string amplitudes of this superstring action and we
obtain the corresponding Koba-Nielsen formula of the well known
amplitudes in the NSR formalism
\cite{GSWSuperstrings,ItoyamaSuperAmps}. This is carried out
explicitly by performing the path integration of this superstring
action with $N$ tachyonic vertex operators in the spirit of Refs. \cite{Zabrodin,GhoshalBfield}. 
We obtain the amplitudes previously found in Refs.
\cite{Arefeva:1988qi,MarshakovNew}.

The article is organized as follows: in Section \ref{Secp-Fermions} we introduce the $p$-adic fermionic action and give
the details of its construction. Later we study its Green's function
in the $p$-adic Fourier representation. Section \ref{SecSuperAction} is devoted to
studying the superstring action. We prove that it satisfies
a supersymmetric invariance with the appropriate tools of the
$p$-adic construction. The action is also written in terms of $p$-adic superfields. 
In Section \ref{SecTreeAmps} we
carry out the explicit computations of the tree-level amplitudes
through correlation functions of $N$ vertex operators. We use the path integration of this theory
to give a Koba-Nielsen type formula for this amplitude. In Section \ref{Sec4-Amps} the 4-point
amplitudes are calculated explicitly and are compared
with those obtained in Refs. \cite{Arefeva:1988qi,MarshakovNew}. Section \ref{Remarks}
contains the final discussions. Finally, in appendix \ref{p-AdicStuff}, we give a brief overview of the $p$-adic tools in order to introduce the notation and conventions that we will follow in the present article. In appendices \ref{AppVertex} and \ref{AppDerivative} we discuss the vertex operators used and
a brief discussion of the fermionic propagator is also included, respectively.

\section{A $p$-adic fermionic action}
\label{Secp-Fermions}

In this section, we consider a $p$-adic analogue of a fermionic action corresponding to the fermionic sector of the Archimedean worldsheet superstring action in the superconformal gauge. The action has previously appeared in \cite{MarshakovNew,GubserMelonic}, but a thorough analysis of its properties was not developed. The proposed action is the following
\be
S_F[\psi]=\frac{\st(-1)p}{2}\frac{1}{\alpha'}\int_{\qp^2}\psi^\mu(x)\eta_{\mu\nu}\frac{\st(x-y)}{|x-y|_p^{s+1}}\psi^\nu(y)dydx,
\label{SF} \ee 
where $\eta_{\mu \nu}=$diag$(-1,1,\dots,1)$ is the Minkowski metric,  $\psi:\qp \to \Lambda$, is a `worldsheet' $p$-adic field valued  in the Grassmann number field $\Lambda$, and $\st$ is the $p$-adic sign function\footnote{Basically there are $3$ distinct non-trivial sign functions determined by $\tau\in \{\epsilon,p,\epsilon p\}$ ($\epsilon$ is a $(p-1)$-rooth of unity). One doesn't always have $\st(-1)=-1$, such requirement implies the restrictions $\tau\neq \epsilon$ and $p \equiv 3\ \mathrm{mod}\ 4$. See the appendix \ref{p-AdicStuff}. }. Some comments regarding this proposal are in order. The action \eqref{SF} requires $\st(-1)=-1$, otherwise it will vanish identically due to the anti-commutativity of $\psi$. The authors of \cite{QuGaoSpinor,Gubser:2018cha} considered a complex $\psi$,
that would imply having  $\psi^*$ instead of one of the fields $\psi$ in \eqref{SF}. This eliminates the
need for including $\st$ at all, but whether or not $\st(-1)=-1$ determines if \eqref{SF} is
symmetric or antisymmetric under the exchange $\psi \leftrightarrow
\psi^*$. We keep the fields real as they are closer to the Archimedean case.

The action \eqref{SF} is closely related to a \textit{twisted} Vladimirov derivative recently studied in \cite{WZG,DuttaGhoshal:2020}. It is straightforward to show that \ba \int_{\qp}\psi^\mu(x)D_s^\tau\psi^\nu(x) dx \nonumber
+ \int_{\qp}\psi^\nu(x)D_s^\tau\psi^\mu(x) dx \\
=(1-\st(-1))\int_{\qp^2}\psi^\mu(x)\frac{\st(x-y)}{|x-y|_p^{s+1}}\psi^\nu(y)dydx,
    \label{SFIdentity}
\ea where $D_s^\tau$ is the generalized or twisted Vladimirov derivative defined by
\be D_s^\tau\psi^\mu(x) := \int_{\qp}
\frac{\psi^\mu(y)-\psi^\mu(x)}{\st(x-y)|x-y|_p^{s+1}}dy. \label{Vderivative}\ee Notice
the two terms on the first line of \eqref{SFIdentity} differ only by
the exchange of indices $\mu \leftrightarrow \nu$. Thus when we
contract with the metric $\eta_{\mu\nu}$ they become equal and we
have \be
\int_{\qp}\eta_{\mu\nu}\psi^\mu(x)D_s^\tau\psi^\nu(x) dx
=\frac{1-\st(-1)}{2}\int_{\qp^2}\psi^\mu(x)\eta_{\mu\nu}\frac{\st(x-y)}{|x-y|_p^{s+1}}\psi^\nu(y)dydx.\\
    \label{SFIdentity2}
\ee We can see then that the fermionic action is almost the same as
the bosonic action, except for the inclusion of the sign function and the parameter $s$. To connect with
previous work and the Archimedean case, we will eventually make
$s=0$, but for now we leave it general.

Unfortunately, with the inclusion of $\st$, there is no value of $s$
for which \eqref{SF} is conformally invariant (invariant under
PGL$(2,\qp)$ transformations), although we do have translation invariance. Later in section \ref{AmpInts} we will implement conformal symmetry directly in the definition of the amplitudes.

\subsection{Fermionic Green's function}
Action \eqref{SF} may be rewritten in a simpler quadratic form. Defining a suitable operator $\Delta^\tau_s$, the action is proportional to $\psi\cdot \Delta^\tau_s \psi $. The purpose of this section is to obtain the inverse operator of $\Delta^\tau_s$. The computation is done using Fourier analysis, this is close in spirit to the computation done in \cite{GhoshalBfield} for bosonic strings in an external $B$-field. A more rigorous study of Green's functions for simple Vladimirov derivatives can be found in \cite{YauGreens}.\\
First, we define the function \be
\F_{\mu\nu}^\fer(s;x-y)=\eta_{\mu\nu}\F^\fer_s(x-y)=\eta_{\mu\nu}\frac{\st(x-y)}{|x-y|_p^{s+1}}.
\ee The superscript $F$ means we are working with the fermionic
sector. $\F^\fer_s$ can be regarded as  the integration kernel for the
operator. Equivalently, we define operator
$\Delta^\tau_s$ acting as the convolution with the function
$\F^\fer_s(\cdot)$, \be \Delta^\tau_s \psi^\mu (x)=(\F^\fer_s\ast
\psi^\mu)(x). \ee We also define
$\G^{\mu\nu}_\fer(s;x-y)=\eta^{\mu\nu}\G_\fer^s(x-y)$ as the inverse of
$\F^\fer_s$, such that \be \int_{\qp} \F^\fer_s(x-z) \G_\fer^s(z-y) dz =
\delta(x-y). \label{GreensDef} \ee 
In Fourier space (see \ref{AppFourier}), this equation reads
 \be \widetilde{\G}_\fer^s(\omega)=\frac{1}{\widetilde{\F}^\fer_s
(\omega)}. \ee To obtain $\G_\fer^s$ we first calculate the Fourier
transform of $\F^\fer_s$ 
$$
\widetilde{F}^\fer_s (\omega)=\pint \chi (\omega x)
\frac{\st(x)}{|x|_p^{s+1}}dx $$ \be = \begin{cases}
       |\omega|_p^s \st(\omega)L(\tau,p) p^{-s-1}, & \tau\neq \epsilon \\
       |\omega|_p^{s}\st(\omega)p^{-s}\frac{1+p^{-s-1}}{p^{-s}+1}, & \tau= \epsilon
   \end{cases} ; \quad {\rm Re}(s)<0.
   \label{FourierTF}
\ee Here $L(\tau,p)=\st(p)\sum_{a=1}^{p-1}\st(a)\chi(a/p)$, but the
exact value is not relevant. Then we have
$$
{\G}^\fer_s (x-y)=\pint \chi (-\omega(x-y))
\frac{\st(\omega)}{|\omega|_p^{s}\mathcal{C}(s,\tau)}d\omega
$$
\be
 =
\begin{cases}    \st(-1)p|x-y|_p^{s-1} \st(x-y),
  &\tau\neq \epsilon \\  \st(-1)p|x-y|_p^{s-1} \st(x-y)\frac{(1+p^{-s})^2}{(1+p^{-s-1})(1+p^{-s+1})}, & \tau= \epsilon \end{cases}; \quad
  {\rm Re}(s)<0,
\label{GreensFunction}\ee where $\mathcal{C}(s,\tau)$ is the coefficient of $
|\omega|_p^{s}\st(\omega)$ in Eq. \eqref{FourierTF}. We are interested in the case $\tau\neq \epsilon$. We compensate for the extra factor $\st(-1)p$ by adding it as a coefficient in front
of the action \eqref{SF}.
In appendix \ref{AppDerivative} we show explicitly that, as usual, the Green's function is the same as the fermion field two-point function.

\section{A `worldsheet' action for the $p$-adic superstring}
\label{SecSuperAction}

In this section we propose a prospect of a non-Archimedean superstring action. This action can be compared with the usual
Archimedean superstring action in the superconformal gauge. It is also shown that this action satisfies 
a supersymmetric transformation in the $p$-adic context. Moreover a superspace formulation is also provided.    

We now propose an action $I_{S}[X,\psi]= I_B[X] + I_F[\psi]$ that
describes the non-Archimedean superstring, and given by the sum of a
bosonic action $I_B[X]$ expressed as \be I_B[X]=
\frac{T_0}{2}\int_{\qp^2}\eta_{\mu\nu}\frac{(X^\mu(x)-X^\mu(y))(X^\nu(y)-X^\nu(x))}{|x-y|_p^2}dydx,
\ee 
and a fermionic action $I_F[\psi]$ given by Eq. (\ref{SF}). Thus
$I_{S}[X,\psi]$ is written as
$$
I_{S}[X,\psi]
=\frac{T_0}{2}\int_{\qp^2}\eta_{\mu\nu}\frac{(X^\mu(x)-X^\mu(y))(X^\nu(y)-X^\nu(x))}{|x-y|_p^2}dydx
$$
\be +
\frac{\st(-1)p}{2\alpha'}\int_{\qp^2}\psi^\mu(x)\eta_{\mu\nu}\frac{\st(x-y)}{|x-y|_p^{s+1}}\psi^\nu(y)dydx.
\ee This action can be rewritten as \be I_{S}=
-T_0\int_{\qp}\eta_{\mu\nu}X^\mu(x)[D_{1}^1X^\nu](x)dx +
\frac{\st(-1)p}{2\alpha'}\int_{\qp}\eta_{\mu\nu}\psi^\mu(x)[\Delta^\tau_s\psi^\nu](x)dx,
\ee where $D_1^1$ is the Vladimirov derivative given in Eq. (\ref{Vderivative}) with $s=1$ and $\tau =1$, and $T_0=\frac{p(p-1)}{4(p+1)\ln p}\frac{1}{\alpha'}$, (see
\cite{Zabrodin,Spokoiny}).
	
\subsection{Equations of motion}
Now we compute the variation of the action for the bosonic and fermionic
fields. Because both fields are real valued, the variation is the
same as in the usual Archimedean case. It is given by
$$
I_B[X+\delta X]=-T_0\pint \eta_{\mu\nu}(X^\mu + \delta X^\mu)(D_1^1
X^\nu + D_1^1 \delta X^\nu)
$$
\be =I_B[X] - 2T_0 \pint \eta _{\mu\nu}\delta X^\mu [D_1^1 X^{\nu}]
+ \mathcal{O}(\delta X^2). \ee This implies that \be \delta
I_B[X]=-2T_0\pint \eta _{\mu\nu}\delta X^\mu(x) [D_1^1 X^{\nu}](x)
dx. \label{BosVar} \ee
In the previous computation we used the following fact 
\be \pint f(x) [D_1^1g](x)dx=\pint [D_1^1f](x)g(x)dx, \ee which
assumes Fubini's theorem, $|-1|_p=1$ and a symmetric
$\eta_{\mu\nu}$. Similarly for the fermionic action we
have
$$
I_F[\psi + \delta \psi] =
\frac{\st(-1)p}{2\a}\int_{\qp^2}\eta_{\mu\nu}\Big(\psi^\mu(x) +
\delta\psi^\mu(x)\Big)\frac{\st(x-y)}{|x-y|_p^{s+1}}\Big(\psi^\nu(y)
+ \delta\psi^\nu(y)\Big)dxdy
$$
\be =S_F[\psi] +
\frac{\st(-1)p}{2\alpha'}\int_{\qp^2}\eta_{\mu\nu}\frac{\st(x-y)}{|x-y|_p^{s+1}}\Big(\delta\psi^\mu(x)\psi^\nu(y)
+ \psi^\mu(x)\delta\psi^\nu(y)\Big)dxdy + \mathcal{O}(\delta
\psi^2).
\ee
This implies that\footnote{We could have started with the equivalent definition
of $S_F\sim \psi \cdot D_s^\tau \psi$. If one does this, one would
eventually get in the integrand $\delta \psi \cdot D_s^\tau \psi +
D_s^\tau \psi \cdot \delta\psi + (1-\st(-1))\delta \psi \cdot
\Delta^\tau_s \psi $. The first two terms cancel and we end up with
the same result as in \eqref{FerVar}. Notice that this is analogous to a
boundary term after integration by parts.}
$$
\delta
I_F[\psi]=\frac{\st(-1)-1}{2\alpha'}p\int_{\qp^2}\eta_{\mu\nu}\frac{\st(x-y)}{|x-y|_p^{s+1}}\delta\psi^\mu(x)\psi^\nu(y)dxdy
$$
\be =\frac{\st(-1)-1}{2\alpha'}p\pint
\delta\psi^\mu(x)[\Delta_{s,\mu\nu}^\tau \psi^{\nu}](x)dx.
\label{FerVar} \ee These two variations imply the following equations of
motion for the bosonic and fermionic fields, $X$ and $\psi$
\be -2T_0\eta_{\mu\nu}[D_1^1X^\nu](x)=0, \ \ \ \ \
 \frac{\st(-1)-1}{2\alpha'}p[\Delta_{s,\mu\nu}^\tau \psi^{\nu}](x)=0. \ee

\subsection{Supersymmetry transformation}
We can see that the proposed action has associated an infinitesimal supersymmetric transformation. 
From \eqref{FerVar} and \eqref{BosVar}, the variation of the `worldsheet' action $I_{S}[X,\psi]$ is written as
\small \be \delta I_{S}[X,\psi] = -2T_0\pint \eta _{\mu\nu}\delta
X^\mu(x) [D_1^1 X^{\nu}](x) dx + \frac{\st(-1)-1}{2\alpha'}p\pint
\delta\psi^\mu(x)[\Delta_{s,\mu\nu}^\tau \psi^{\nu}](x)dx. \ee
\normalsize Let us insert the following variations: \be \delta X^\mu
= A\lambda \Delta^\tau_s \psi^\mu, \quad \quad \delta \psi ^\mu =
B\lambda D_1^1 X^\mu, \ee where $A$ and $B$ are quantities to determine and $\lambda$ is a Grassmann
parameter of the transformation. Then we have \small
$$
\delta I_{S}[X,\psi] = \pint \eta _{\mu\nu}\bigg(-2T_0 A\lambda
[\Delta^\tau_s \psi^\mu](x)[D_1^1 X^{\nu}](x) +
\frac{\st(-1)-1}{2\alpha'}pB\lambda [D_1^1 X^\mu](x)[\Delta^\tau_s
\psi^{\nu}](x)\bigg)dx
$$
\normalsize \be =\pint  \left(\lambda[\Delta^\tau_s \psi]\cdot [D_1^1
X]\right)\left( -2T_0 A + \frac{\st(-1)-1}{2\alpha'}pB \right)dx, \ee where
we can see that choosing constants $A$ and $B$ such that
$-2T_0 A + \frac{\st(-1)-1}{2\alpha'}pB=0$, we will get $\delta
I_{S}[X,\psi]=0$. Then for example, when $\st(-1)=-1$, the
transformation \be \delta X^\mu = \lambda \Delta^\tau_s \psi^\mu;
\quad \quad \delta \psi ^\mu = -\frac{2\alpha'}{p} T_0 \lambda D_1^1 X^\mu, \ee is an
infinitesimal supersymmetric transformation of $I_{S}[X,\psi]$.

\subsection{Superspace description of the $p$-adic superstring}
It is desirable to have a more efficient and concise approach to describe the $p$-adic superstring such as the superspace approach. In \cite{Dragovich:2005we} a notion of superspace over $\qp$ is introduced. Motivated by this, we follow \cite{ItoyamaSuperAmps} and define a superfield and a
superoperator for our model. After some work one can find that the
superoperator and the superfield are of the following form
$$
\mathcal{X}^\mu(x,\theta) = AX^\mu(x) + B\theta \psi^{\mu} (x),
$$
\be \mathcal{D}_s^\tau =  a\theta D_1^1 + b\Delta_s^\tau \partial_\theta.
\label{SuperPreDefs} \ee Then \ba \mathcal{D}_s^\tau \mathcal{X}^\mu
&=& aA\theta D_1^1 X^\mu + bB \Delta^\tau_s \psi ^\mu,\nn \\
\mathcal{X} \cdot \mathcal{D}_s^\tau \mathcal{X} &=& aA^2 \theta
X\cdot D_1^1 X + bABX \cdot \Delta_s^\tau \psi  + bB^2 \psi \cdot
\Delta_s^\tau \psi. \ea In order to write down the action in the
following form \be I_{S}[X,\psi]=\int\limits_{\qp} \int d\theta\
\eta_{\mu\nu}\mathcal{X}^\mu [\mathcal{D}_s^\tau \mathcal{X}^\mu]
dx, \label{SuperAction} \ee  we need to choose the constants $a$,
$b$, $A$ and $B$, such that $aA^2=-T_0$ and $bB^2=\st(-1)p/{2\alpha'}$.
This is underdetermined, however, there is a further constraint that
can be considered. We must also require to have the vetex operators given by
\be {\cal V}(k_\ell;y_\ell)=\int d\theta_\ell
e^{ik_\ell\cdot \mathcal{X}_\ell(y_\ell)}. \label{SuperVertex} \ee A
comparison with Eq. \eqref{Vxs} (See below) shows that we need $A=1$ and $B=-i$.
This determines $a=-T_0$ and $b=-\st(-1)p/2\alpha'$. Thus the appropriate
choice of constants in Eq. \eqref{SuperPreDefs} to obtain
\eqref{SuperAction} and \eqref{SuperVertex} is \ba
\mathcal{X}^\mu(x,\theta) &=& X^\mu(x) - i\theta \psi^{\mu} (x),\\
\mathcal{D}_s^\tau &=& -T_0 \theta D_1^1 - \frac{\st(-1)p}{2\alpha'}
\Delta_s^\tau \partial_\theta. \ea The Green's function of the differential operator $\mathcal{D}$
is given by
$$
\mathcal{G}_s(x-y;\theta,\theta')=\frac{\alpha'}{1-s}\ln\left(|x-y|_p^{1-s}+\st(x-y)(1-s)\theta\theta'\right)
$$
\be = \alpha' \ln |x-y|_p +
\alpha'\theta\theta'\frac{\st(x-y)}{|x-y|_p^{1-s}}, \ee and satisfies \be
\pint \mathcal{D}_s^\tau (x-z;\theta)
\mathcal{G}_s(z-y;\theta,\theta') dz =\delta(x-y)(\theta-\theta'). \ee

\section{Tree-level amplitudes of the $p$-adic superstring }
\label{SecTreeAmps}
In this section we obtain the tree-level amplitudes of our $p$-adic superstring model. They are obtained through the computation of correlation functions of vertex operators.

The $N$-point function for this system is given by the insertion of
$N$ vertex operators of the form
\be
{\cal V}(y_\ell)=k_\ell\cdot \psi(y_\ell) e^{ik_\ell \cdot
X(y_\ell)} =\int d\theta_\ell e^{ik_\ell\cdot X(y_\ell) +
\theta_\ell k_\ell\cdot \psi(y_\ell)}, \label{Vxs}
\ee
where $y_\ell$ is the insertion point, and $\theta_l$ are auxiliary Grassmann variables. Inserting vertex
operators inside the path integral is equivalent to having a generating function with appropriately chosen sources for both fermions and
bosons. The integration of the bosonic part can be obtained in a
standard way using the corresponding Green's function
$\G^\bos_{\mu\nu}(x-y)=-\alpha' \log|x-y|_p$ \cite{GhoshalBfield}.
Here we will perform the analogous computation carried out in \cite{GhoshalBfield}, but now for the fermionic sector.

We start by recalling the operator $\Delta^\tau_{\mu\nu}$ (from now on we will omit the explicit dependance on the parameter $s$), and define its inverse
$(\Delta_\tau^{-1})_{\mu\nu}$ by \be
[\Delta_{\mu\nu}^\tau K^\nu](x)=(\F_{\mu\nu}^\fer \ast
K^\nu)(x)=\int_{\qp} \F_{\mu\nu}^\fer (x-y) K^\nu(y) dy \ee and \be
[(\Delta^{-1}_\tau)_{\mu\nu}K^\nu](x)= (\G_{\mu\nu}^\fer \ast
K^\nu)(x)=\int_{\qp} \G_{\mu\nu}^\fer (x-y) K^\nu(y) dy. \ee
Using Fubini's theorem it is straightforward to check that
$$
((\Delta^{-1}_\tau)^{\mu\alpha}[\Delta_{\alpha\nu}^\tau K^\nu])(x)
=K^\mu(x);
$$
$$
(\Delta_{\mu\alpha}^\tau[(\Delta^{-1}_\tau)^{\alpha\nu}K_\nu])(x)
=K_\mu(x).%
$$
Now notice the following identity for general functions $f$ and $g$
$$
\pint f(x)[\Delta^\tau_{\mu\nu}g](x) dx
=\int_{\qp ^2}f(x)\F_{\mu\nu}^\fer(x-y)g(y)dydx
$$
$$
=\pint\st(-1)\left[ \pint f(x)\F_{\mu\nu}^\fer(y-x)dx\right]g(y) dy
$$
$$
=\st(-1)\pint [\Delta^\tau_{\mu\nu}f](x)g(x)dx,
$$
where we used $\F^F_{\mu\nu}(x-y) =\st(-1)\F^F_{\mu\nu}(y-x)$. In words, the operator $\Delta^\tau_{\mu\nu}$ inside the integral may switch its action to the rest of the integrand at the cost of a sign.  From this we can easily get that
\be
\pint [(\Delta^{-1}_\tau)^{\nu\alpha}K_\alpha](x)
[\Delta_{\nu\beta}^\tau \psi^\beta](x) dx
 =\st(-1)\pint K^\alpha(x) \psi_\alpha (x) dx.	\label{Subtlesgn}
 \ee
With these results one can verify the following identity
$$
-\frac{1}{2}\int_{\qp} \psi^\mu(x)[\Delta_{\mu\nu}^\tau\psi^\nu](x)dx +
\int_{\qp} K_\mu(x) \psi^\mu (x) dx
$$
$$
 = -\frac{1}{2}\int_{\qp}
\bigg(\psi^\mu(x) -\st(-1) [(\Delta^{-1})^{\mu\alpha}_\tau
K_\alpha](x)\bigg) \bigg[\Delta_{\mu\nu}^\tau \bigg(\psi^\nu
-\st(-1) [(\Delta^{-1})^{\nu\beta}_\tau K_\beta] \bigg)\bigg](x) dx
$$
\be
 +\frac{1}{2}\int_{\qp}K^\mu (x)[(\Delta^{-1})^{\mu\nu}_\tau
K_\nu](x) dx, \ee that is the fermion analogue of the well known
relation  $-\frac{1}{2}x^T \cdot A \cdot x + K^T \cdot x =
-\frac{1}{2}(x^T-K^T \cdot A^{-1}) \cdot A \cdot (x-A^{-1}\cdot K) +
\frac{1}{2} K^T \cdot A^{-1}\cdot K$ for the bosonic finite
dimensional case where $A$ is a matrix, $x$ and $K$ are column
vectors and $T$ denotes the transpose operation. Thus we are able to obtain the generating function from the path integral with bosonic sources $J^\mu$ and fermionic sources $K^\mu$ in this $p$-adic setting \small
$$
\mathcal{Z}[J,K]=\frac{1}{\mathcal{Z}}\int D\psi \int DX
\exp\left\{-I_B[X] - I_F[\psi]+\int_{\qp}J_\mu (x) X^\mu (x) dx +
\int_{\qp}K_\mu (x) \psi^\mu (x) dx\right\}
$$
\be
 =\exp\left\{
\frac{1}{2}\int_{\qp^2} J^\mu (x) \G^\bos_{\mu\nu}(x-y)J^\nu (y)
dxdy + \frac{\a}{2\st(-1)p}\int_{\qp^2} K^\mu (x) \G^\fer_{\mu\nu} (x-y)K^\nu
(y) dxdy \right\}. \label{GenFunction}\ee
\normalsize This generating function is equivalent to the $N$-point
function if we use the sources \be \mathcal{J}^\mu (x) =
i\sum_{l=1}^N k_l^\mu \delta (x-y_l), \quad \mathcal{K}^\mu (x) =
\sum_{m=1}^N \theta_m k_m^\mu \delta (x-y_m), \ee and integrate out the $\theta_m$ variables such that \be \left\langle
V(y_1)\cdots V(y_N)\right\rangle =\int d\theta_1 \dots d\theta_N
\mathcal{Z}[\mathcal{J},\mathcal{K}]. \label{ExpVal} \ee
The dependence on the insertion points is left implicit in the sources on the right hand side. This is the analogue for the usual basic prescription to obtain $N$-point amplitudes. It is our starting point in order to get explicit expressions for the tree-level amplitudes. 
\subsection{Integral expression} \label{AmpInts}
The next step is to obtain a clearer and more explicit expression of amplitudes \eqref{ExpVal}. As mentioned in section \ref{Secp-Fermions}, we kept $\st(-1)$ general to exhibit the subtleties that come with it (see Eq. \eqref{Subtlesgn}), and so that the previous expressions can also be used if one were to consider complex valued Grassmann fields, for which the action \eqref{SF} doesn't identically vanish. However in what follows and for the rest of the paper we will take $\st(-1)=-1$, otherwise all of the following computations would give identically $0$.\\
First we obtain
$$
\frac{1}{2}\int_{\qp^2} \mathcal{K}^\mu (x) \G^\fer_{\mu\nu}
(x-y)\mathcal{K}^\nu (y) dxdy
=\frac{1}{2}\sum_{m,n=1}^N \theta_m \theta_n k_m^\mu k_n^\nu \int_{\qp^2} \delta(x-y_m) \G^\fer_{\mu\nu} (x-z)\delta(z-y_n) dxdz\\
$$
$$
=\frac{1}{2}\sum_{\substack{m,n=1\\ m\neq n}}^N \theta_m \theta_n
k_m^\mu k_n^\nu\G^\fer_{\mu\nu} (y_m-y_n)
$$
$$
=\frac{1}{2}\sum_{ m< n}k_m \cdot k_n \left[\theta_m \theta_n \G^\fer (y_m-y_n)+\theta_n \theta_m \G^\fer (y_n-y_m)\right]  \label{DoubleSum}\\
$$
\be
=\sum_{ m< n}k_m \cdot k_n \theta_m \theta_n
\G^\fer (y_m-y_n)
=\st(-1)p\sum_{ m< n}\theta_m \theta_n \ k_m \cdot
k_n \frac{\st(y_m-y_n)}{|y_m-y_n|_p^{1-s}}. \ee
Similarly\footnote{The astute reader may have noticed that we
seemingly forgot the case $m=n$ in the sum. A careful calculation
shows that those terms are vanishing. The subtlety lies in the fact that in
order to have a well defined integration of the Green's functions,
we must integrate over the $p$-adic plane such that $x\neq y$. This
is because $\G^\bos$ is singular when its argument is 0, and $\G^\fer$ is ill-defined
at 0 ($\st(0)$ is ill-defined). Then when we integrate one of the
deltas in the first line of \eqref{DoubleSum}, we are left with
something proportional to $\delta(x-y_n)\Big\rvert_{x\neq y_n}$,
which is always 0.} \be \frac{1}{2}\int_{\qp^2} \mathcal{J}^\mu (x)
\G^\bos_{\mu\nu}(x-y)\mathcal{J}^\nu (y) dxdy =\alpha'\sum_{ m<n}k_m
\cdot k_n\log|y_m-y_n|_p. \ee Using this and the properties of Grassmann variables (see Eq. \eqref{LogSum} in Appendix B) we can write
$$
\mathcal{Z}[\mathcal{J},\mathcal{K}]=\exp\left\{ \sum_{m<n}\alpha'k_m\cdot
k_n \left[ \log|y_m-y_n|_p +
\theta_m\theta_n\frac{\st(y_m-y_n)}{|y_m-y_n|_p^{1-s}}\right]
\right\}\nn
$$
$$
= \exp\left\{ \sum_{m<n}\frac{\alpha'}{1-s}k_m\cdot k_n \left[
\log\left(|y_m-y_n|_p^{1-s} +
\st(y_m-y_n)(1-s)\theta_m\theta_n\right)\right]
\right\} 
$$
\be
 =\prod_{m<n} \left(|y_m-y_n|_p^{1-s} +
\st(y_m-y_n)(1-s)\theta_m\theta_n\right)^{k_m\cdot k_n
\frac{\alpha'}{1-s}}.\label{ZJK} \ee We will also make now $s=0$ for the rest of the article\footnote{Again, a careful reader might doubt about the validity of this value for $s$, since in \eqref{GreensFunction} we determined that $\text{Re}(s)<0$. This is common in physics, when we set $s=0$, we mean to take the analytic continuation of the result \eqref{GreensFunction}.}. This is just to make it closer to the Archimedean case and avoid carrying the parameter around. Then finally we define the $N$-point amplitudes as follows
$$
\mathcal{A}^{(N)}_p
(\mathbf{k}) :=\mathcal{N}\int_{\qp^N} \left\langle V(k_1;y_1)\cdots
V(k_N;y_N)\right\rangle \prod_{m<n}\st(y_m-y_n)\prod_{i=1}^N dy_i
$$
\be
 =\mathcal{N}\int_{\qp^N} \int
\prod_{j=1}^N d\theta_j \prod_{m<n} \left(|y_m-y_n|_p +
\st(y_m-y_n)\theta_m\theta_n\right)^{\alpha'k_m\cdot
k_n}\st(y_m-y_n)\prod_{i=1}^N dy_i , \label{SuperAmps} \ee where $\mathcal{N}$ is a normalization constant. This is the analogous result of the usual 
Archimedean result. The factor $\prod_{m<n}\st(y_m-y_n)$ is added to implement conformal invariance. These amplitudes were also proposed in \cite{Arefeva:1988qi,MarshakovNew}. As we can see it is quite similar to the usual result \cite{GSWSuperstrings}, the only difference is the appearance of the sign functions.

\subsection{Integrating the Grassmann variables}
The purpose of this subsection is to get the tree-level open string
amplitudes by carrying out integration \eqref{SuperAmps}. If we go back
to the first line of \eqref{ZJK}, we can factorize the exponential and expand the
fermionic part \be \exp\left\{ \sum_{m<n}\a k_m\cdot
k_n\frac{\st(y_m-y_n)}{|y_m-y_n|_p} \theta_m\theta_n  \right\}
=\prod_{m<n}\left(1+\a k_m\cdot
k_n\frac{\st(y_m-y_n)}{|y_m-y_n|_p}  \theta_m\theta_n \right). \label{FermExpand}
\ee Notice that the terms will always have an even number of
$\theta_n$ variables. Considering that there are $N$ Grassmann integrations, we
conclude that the only nonvanishing amplitudes are for even
insertions. From now on we consider to have an even $N$. Remember that $\theta_m ^2 =0$ and terms with less than
$N$ $\theta$s will be annihilated by the integrals. Therefore only
terms with $N$ distinct $\theta_n$ variables survive.

All of these amounts to the amplitudes being composed of $(N-1)!!$ terms, this is Wick's theorem.
The terms differ in specific permutations of the $N$ insertion points. We will define these permutations a few lines below, but for now consider them of the form  \be
\theta_{m_1}\theta_{n_1}\cdots\theta_{m_{N/2}}\theta_{n_{N/2}},
\quad \quad m_i<n_i, \ m_i\neq m_j,\ n_i\neq n_j. \label{Perms} \ee
Integrating the $\theta_m$ variables in \eqref{FermExpand}, we are left with 
\be
\a \sum_{P}(-)^{P}\prod_{i=1}^{N/2} k_{P(2i-1)}\cdot
k_{P(2i)}\frac{\st(y_{P(2i-1)}-y_{P(2i)})}{|y_{P(2i-1)}-y_{P(2i)}|_p},
\label{FermionSum} \ee

\noindent where $P$ are permutations of the form \eqref{Perms} and $(-)^P$ is its sign. With
this the amplitudes \eqref{SuperAmps} can now be written in the
Koba-Nielsen form \[ \mathcal{A}^{(N)}_p (\mathbf{k})
=\mathcal{N}\int_{\qp^N}\prod_{m<n}|y_m-y_n|_p^{\alpha' k_m\cdot k_n}\st(y_m-y_n)\]
\be \times \sum_{P}(-)^{P}\prod_{i=1}^{N/2} k_{P(2i-1)}\cdot
k_{P(2i)}\frac{\st(y_{P(2i-1)}-y_{P(2i)})}{|y_{P(2i-1)}-y_{P(2i)}|_p}
\prod_{i=1}^N dy_i. \label{SuperAmpsSum} \ee  This is made
cleaner by defining the following amplitude 
\[ A^{(N)}_p(\mathbf{k},P_N) :=\mathcal{N}\prod_{m<n}(k_m\cdot k_n)^{q_{mn}(P_N)}\]
\be \times \int_{\qp^N} \prod_{m<n}|y_m-y_n|_p^{\alpha' k_m\cdot k_n
-q_{mn}(P_N)}[\st(y_m-y_n)]^{1+q_{mn}(P_N)} \prod_{i=1}^N dy_i, \label{CleanerAmplitude} \ee where $P_N$ is a
general permutation of $N$ elements and $q_{mn}(P_N)$ is defined as
follows
$$
q_{mn}(P):= \begin{cases}
1 , \, \exists\ i\in \{1,\dots,N/2\} ; \ \ \ (m,n)=(P(2i-1),P(2i))    \\
0 ,\,   \text{otherwise}
            \end{cases}
$$
\be = \sum_{i=1}^{N/2} \delta_m^{P(2i-1)}\delta_n^{P(2i)},\label{qmn} \ee
notice that $q_{mn}\neq q_{nm}$. It is now a good time to define more rigorously the
permutations appearing in the sum in \eqref{SuperAmpsSum}. We set
$\widetilde{P}$ as the set of permutations of $N$ elements such that
for every $i\in \{1,\dots,N/2\}$ we have
$\widetilde{P}(2i-1)<\widetilde{P}(2i)$ and
$\widetilde{P}\equiv\widetilde{P}'$ if
$\widetilde{P}(2i-1)=\widetilde{P}'(2j-1),
\widetilde{P}(2i)=\widetilde{P}'(2j)$ for $1\leq j\leq N/2$. (All permutations that differ by the exchange of any two pairs of consecutive elements are equivalent.) Notice that we have in total
$\frac{N!}{(N/2)!2^{N/2}}=(N-1)!!$ elements in $\widetilde{P}$, as we should. Now
the amplitudes have a more elegant and deceivingly concise form \be \mathcal{A}^{(N)}_p (\mathbf{k})
=\sum_{\widetilde{P}}(-)^{\widetilde{P}}A^{(N)}_p(\mathbf{k},\widetilde{P}).
\label{ElegantA} \ee

\subsubsection{Conformal Symmetry}\label{Symms}

In this section, we proceed to check the conformal invariance of the amplitudes \eqref{ElegantA}. In the usual case, the gauge fixing of the symmetries of worldsheet diffeomorphisms and Weyl transformations can be carried out. However, it is not completely fixed and there is a remnant symmetry on the two-sphere, this is the PSL$(2,\mathbb{C})$ symmetry \cite{GSWSuperstrings}. In the present case, even though 
the action does not have the conformal symmetry, we will find the conditions under which it can be implemented at the level of the amplitudes. These conditions involved the sign functions. This procedure works for the $p$-adic bosonic string where $k^2=2/\a$ \cite{Brekke:1993gf}. In our case we have $k^2=1/\a$, however, the factors $|y_m-y_n|_p^{-1}$ coming from the fermionic sector described above save the day. Something similar will happen to the $\st$ functions. If you accept that this procedure can be done for our case, go ahead to the next section.\\
Consider the transformation $y_m=\frac{a\bar{y}_m+b}{c\bar{y}_m+d}; \text{with }ad-cb=1$ for the integrand in \eqref{CleanerAmplitude} with a given permutation $\widetilde P$, then we have \be
|y_m-y_n|_p^{s_{mn}}=|\bar{y}_m-\bar{y}_n|_p^{s_{mn}}|c\bar{y}_m+d|_p^{-s_{mn}}|c\bar{y}_n+d|_p^{-s_{mn}},
\quad dy_m=|c\bar{y}_m+d|_p^{-2}d\bar{y}_m. \label{Mobius1} \ee
Applying the change of variables, we will encounter the following
product
$$
\prod_{m<n}|c\bar{y}_m+d|_p^{-s_{mn}}|c\bar{y}_n+d|_p^{-s_{mn}}
=\prod_{m=1}^{N-1}
|c\bar{y}_m+d|_p^{-\sum_{n>m}s_{mn}}\prod_{n=2}^{N}
|c\bar{y}_n+d|_p^{-\sum_{m<n}s_{mn}}
$$
\be =\prod_{m=1}^{N} |c\bar{y}_m+d|_p^{-\sum_{n\neq m}s_{mn}}.
\label{Prod1} \ee Of course we must set $s_{mn}=\alpha'k_m\cdot
k_n$. We use the momenta conservation $\sum_m k^\mu_m =0$, and that
for the open superstrings $k_m^2=1/\alpha'$,  to obtain \be -\sum_{n\neq
m}s_{mn}=-\alpha'k_m\cdot \sum_{n\neq m} k_n =-\alpha' k_m \cdot
(-k_m) = \alpha' k_m^2 =1. \label{kSum} \ee Then for a fractional
linear transformation the integrand of \eqref{SuperAmpsSum} in the
new variables will have the extra factor \be \prod_{m=1}^N
|c\bar{y}_m +d |_p\prod_{m=1}^N |c\bar{y}_m +d |_p^{-2}\prod_{m=1}^N
|c\bar{y}_m +d |_p =1.\label{ExtraFactor} \ee
Here the first product is a result of \eqref{Prod1} and
\eqref{kSum}, the second product comes from the second equality in
\eqref{Mobius1} and third product is the contribution of the fermionic sector \eqref{FermionSum}\footnote{Had we kept $s$ arbitrary, we would get $\prod_{m=1}^N |c\bar{y}_m +d |_p^{-s}$ on the right side of \eqref{ExtraFactor}.}.\\
We now deal with the sign functions, which is easier. We have
\bs\prod_{m<n}[\st(y_m-y_n)]^{1+q_{mn}}=& \prod_{m<n}[\st(\bar{y}_m-\bar{y}_n)]^{1+q_{mn}}\st(c\bar{y}_m+d)\st(c\bar{y}_n+d)\\
&\times\prod_{m<n} [\st(c\bar{y}_m+d)]^{q_{mn}}[\st(c\bar{y}_n+d)]^{q_{mn}}\\
=\prod_{m<n}[\st(&\bar{y}_m-\bar{y}_n)]^{1+q_{mn}}\prod_{m=1}^{N}[\st(c\bar{y}_m+d)]^{N-1}\prod_{m=1}^{N}\st(c\bar{y}_m+d)\\
=\prod_{m<n}[\st(&\bar{y}_m-\bar{y}_n)]^{1+q_{mn}},
\end{split} \label{sgnProds}\ee
the second product of the third line comes from realizing that for any point $y_a$, the product $\prod_{m<n}\st(c\bar{y}_m+d)$ will give us $[\st(c\bar{y}_a+d)]^{N-a}$. Similarly for $\prod_{m<n}\st(c\bar{y}_n+d)$ we have $[\st(c\bar{y}_a+d)]^{a-1}$. Since $q_{mn}$ is non-zero for each unique pair $mn$, only one such factor appears per point, this explains the last product in the third line. In the last line we used that $N$ is even and $[\st(\cdot)]^2 =1$.\\
With this we have proven the symmetry of the amplitudes. Conformal invariance allows us to fix three insertion points. It is customary to take such points as $0$, $1$, and $\infty$. Here is the explicit transformation that does the job
\[x_i=\frac{(y_{N-1}-y_{N})(y_1-y_{i})}{(y_1-y_{N-1})(y_i-y_{N})}\quad \Leftrightarrow \quad y_i=\frac{x_i\ y_N (y_1 -y_{N-1})+y_1(y_{N-1}-y_N)}{y_{N-1}-y_N+x_i(y_1-y_{N-1})}.\]
This sends $x_1=0,\ x_{N-1}=1,\ x_N=\infty$. We only transform $y_i$ for $i\in \{2,\dots, N-2\}$. We have
\be \begin{split}
y_1-y_i & = \frac{(y_1-y_{N})(y_1-y_{N-1})}{y_{N-1}-y_N+x_i(y_1-y_{N-1})}x_i, \quad
y_i-y_{N-1} = \frac{(y_1-y_{N-1})(y_{N-1}-y_{N})}{y_{N-1}-y_N+x_i(y_1-y_{N-1})}(1-x_i)\\
y_i-y_N &= \frac{(y_1-y_{N})(y_{N-1}-y_{N})}{y_{N-1}-y_N+x_i(y_1-y_{N-1})},\quad
dy_i = \frac{|y_1-y_{N-1}|_p|y_1-y_{N}|_p|y_{n-1}-y_{N}|_p}{|y_{N-1}-y_N+x_i(y_1-y_{N-1})|_p^2}dx_i \\
y_i-y_j &= \frac{(y_1-y_{N-1})(y_1-y_{N})(y_{n-1}-y_{N})}{(y_{N-1}-y_N+x_i(y_1-y_{N-1}))(y_{N-1}-y_N+x_j(y_1-y_{N-1}))}(x_i-x_j). \label{MobiusTransf}
\end{split}\end{equation}
Using these expressions one can check that the integrand factorizes into 
\[dy_1dy_{N-1}dy_N  \prod_{m=2}^{N-2}|x_m|_p^{s_{1m}+q_{1m}}|1-x_m|_p^{s_{m(N-1)}+q_{m(N-1)}}[\st(x_m)]^{1+q_{1m}}[\st(1-x_m)]^{1+q_{m(N-1)}}\]
\[\times \prod_{2\leq m <n\leq N-2}|x_m-x_n|_p^{s_{mn}+q_{mn}}[\st(x_m-x_n)]^{1+q_{mn}} \prod_{m=2}^{N-2}dx_m.\]
We leave the details to the reader. Having $dy_1dy_{N-1}dy_N$ means that we'll have the factor $[\text{Vol}(\qp)]^3$ but this gets canceled after normalizing. Therefore we redefine the amplitudes \eqref{CleanerAmplitude} as
\[ A^{(N)}_p(\mathbf{k},P) =\prod_{m<n}(k_m\cdot k_n)^{q_{mn}}\]
\[\times \int_{\qp^{N-2}} \prod_{m=2}^{N-2}|x_m|_p^{s_{1m}-q_{1m}}|1-x_m|_p^{s_{m(N-1)}-q_{m(N-1)}}[\st(x_m)]^{1+q_{1m}}[\st(1-x_m)]^{1+q_{m(N-1)}}\]
\be\times \prod_{2\leq m <n\leq N-2}|x_m-x_n|_p^{s_{mn}-q_{mn}}[\st(x_m-x_n)]^{1+q_{mn}} \prod_{m=2}^{N-2}dx_m.\label{CleanReducedAmps}\ee

\section{Four-point amplitudes}\label{Sec4-Amps}
As an illustrative example, we show the case $N=4$ in \eqref{ElegantA}. The first ingredients are the permutations $\widetilde{P}$, these are $\widetilde{P}=\{(1234),\ (1324)\ (1423)\}$ with signs $\{1,-1,1\}$. The next step is to determine the non-zero components of $q_{mn}$. They are different depending on the permutation, for example for $(1234)$ only $q_{12}$ and $q_{34}$ are $1$ while the other components equal $0$.\\
Then the amplitude \eqref{ElegantA}, after gauge fixing three points as shown in \eqref{CleanReducedAmps}, is
\[
\mathcal{A}^{(4)}_p
(\mathbf{k}) =(k_{1}\cdot
k_{2})(k_{3}\cdot k_{4})\int_{\qp}|x_2|_p^{\alpha' k_1\cdot k_2-1} |1-x_2|_p^{\alpha' k_2\cdot k_3}\st(1-x_2) dx_2\]
\be
-(k_{1}\cdot k_{3})(k_{2}\cdot k_{4})\int_{\qp}|x_2|_p^{\alpha' k_1\cdot k_2} |1-x_2|_p^{\alpha' k_2\cdot k_3}\st(x_2)\st(1-x_2) dx_2\ee
\[
+(k_{1}\cdot k_{4})(k_{2}\cdot k_{3})\int_{\qp}|x_2|_p^{\alpha' k_1\cdot k_2} |1-x_2|_p^{\alpha' k_2\cdot k_3-1}\st(x_2)dx_2.\]
Being efficient, we do the following integration
$$
\pint |x|_p^{u}|1-x|_p^{v}[\st(x)]^{t_1}[\st(1-x)]^{t_2}dx
=\int_{p\mathbb{Z}_{p}}|x|_p^{u}[\st(x)]^{t_1}dx
$$
$$
+\int_{\mathbb{Z}_{p}^\times}|1-x|_p^{v}[\st(x)]^{t_1}[\st(1-x)]^{t_2}dx
+\int_{\qp\setminus \mathbb{Z}_{p}^\times} |x|_p^{u+v}[\st(x)]^{t_1+t_2}dx
$$
$$
=\int_{p\mathbb{Z}_{p}}|x|_p^{u}[\st(x)]^{t_1}dx
+[\st(-1)]^vp^{-1}\sum_{a=2}^{p-1}[\st(a)]^u[\st(a-1)]^v
$$
$$
+[\st(-1)]^v\int_{p\mathbb{Z}_{p}}|x|_p^{v}[\st(x)]^{t_2}dx
+[\st(-1)]^v\int_{p\mathbb{Z}_{p}}|x|_p^{-u-v-2}[\st(x)]^{t_1+t_2}dx
$$
\be =\begin{cases}
\frac{1-p^{-1}}{p^{u+1}-1} + \frac{1-p^{-1}}{p^{v+1}-1} + \frac{1-p^{-1}}{p^{-u-v-1}-1} +1 - 2p^{-1}, &(t_1,t_2)=(0,0) \\
\frac{1-p^{-1}}{p^{u+1}-1} - p^{-1}=\frac{1-p^u}{p^{u+1}-1}, &(t_1,t_2)=(0,1) \\
\frac{1-p^{-1}}{p^{v+1}-1} - p^{-1}=\frac{1-p^v}{p^{v+1}-1}, &(t_1,t_2)=(1,0) \\
p^{-1}-\frac{1-p^{-1}}{p^{-u-v-1}-1}=\frac{p^{-u-v-2}-1}{p^{-u-v-1}-1} , &(t_1,t_2)=(1,1)
\end{cases}.
\ee
With this result, we can easily see that
$$
\mathcal{A}^{(4)}_p(\mathbf{k})=
(k_1\cdot k_2)(k_3\cdot k_4)\left[\frac{1-p^{\a k_1\cdot k_2-1}}{p^{\a k_1\cdot k_2}-1}\right] 
-(k_1\cdot k_3)(k_2\cdot k_4)\left[\frac{p^{\a k_1\cdot k_3-1}-1}{p^{\a k_1\cdot k_3}-1}\right]
$$
\be
 +(k_1\cdot k_4)(k_2\cdot k_3)\left[\frac{1-p^{\a k_2\cdot k_3-1}}{p^{\a k_2\cdot k_3}-1}\right]
\label{4pAmpInfty}. \ee 
This is the same result reported in \cite{MarshakovNew,Arefeva:1988qi}, where it was computed as a direct analogue of the Archimedean expressions. Comparing to the Archimedean result from \cite{ItoyamaSuperAmps} our amplitudes are very similar in the integral form, and will likely be so for arbitrary points. The main difference is the presence of sign functions, these functions annihilate several terms in the amplitudes when compared to the $p$-adic bosonic string.

\section{Final Remarks}\label{Remarks}

In this article we propose a theory of free $p$-adic worldsheet
superstrings. An
action analogous to the Archimedean case in the superconformal gauge was considered. As usual, the
action consists of two terms, a bosonic and a fermionic part. We
based our proposal in different works that proposed a fermionic
propagator or action. This implies the use of Grassmann valued
$p$-adic fields. To prevent the fermionic term from vanishing
identically, it is necessary to insert an antisymmetric sign
function, {\it i.e.} $\mathrm{sgn}_\tau(-1)=-1$. This restricts the
possible values of $p$ to roughly half the primes, and $\tau$ to 2
of its non-trivial values. We noticed that the fermionic term is in
fact very similar to the bosonic one, the only two differences being
the use of a generalized Vladimirov derivative (that includes
$\mathrm{sgn}_\tau$) and the order of the derivative is decreased by
$1$. From this action we were able to find a supersymmetry
transformation, and write the action in a superspace formalism,
defining a $p$-adic superfield and a derivative superoperator.

Using standard field theory techniques, we obtained the tachyon
$N$-point tree amplitudes. We checked that the fermion propagator is equivalent to the corresponding two-point function. This required a functional derivative for $p$-adic fermion fields. Like in the Archimedean
case, these amplitudes are non-vanishing only for even $N$. A neat
and simple integral form for these amplitudes that is analogous to
the Archimedean case can be given, albeit not very useful for
computations. Explicit results can be obtained by manipulating the
expressions. The procedure is similar to the one of the
Archimedean case. One can see that the amplitudes may be obtained as a
weighted sum of almost purely bosonic amplitudes, except for the
appearance of sign functions in the integrands. Therefore the
amplitudes can be integrated using previously developed techniques
\cite{OurB-Field}. They take the form of rational functions with
momenta variables as powers of $p$. Previous works have shown
that the amplitudes obtained here are integrable and convergent in a
certain region of momenta space \cite{Bocardo-Gaspar:2016zwx}.

Unfortunately, the proposed action is not M\"obius invariant, this
is independently due to both the sign function, and the order of the
Vladimirov derivative in the fermionic term. One may check that the action \eqref{SF} has translation invariance and if $N$ is a multiple of $4$, it is also scale invariant. However this is not sufficient to fix three points.
The guiding principle for $p$-adic theories is the Archimedean
counterpart. In that spirit, we choose to implement conformal invariance (PSL$(2,\mathbb{Q}_p)$
symmetry) by inserting the factor $\prod_{m<n}\st(y_m-y_n)$ in the amplitudes, as shown at the end of section \ref{AmpInts}. With this we can gauge fix three points as usual and greatly reduce computations.
The $4$-point function was obtained explicitly and has crossing symmetry. It coincides with the one obtained in \cite{MarshakovNew}. 

In terms of Beta functions, the amplitudes are very similar to the Archimedean case. This was exploited in \cite{Brekke:1993gf,Arefeva:1988qi} to construct $p$-adic superstring amplitudes.
Compared to the $p$-adic bosonic string, the main difference is the appearance of sign functions, their effect is to eliminate several terms after the integration.
However, it remains unclear whether these amplitudes can come
directly from a Lagrangian. The authors of \cite{GhoshalBfield} first pointed
out in that having $\mathrm{sgn}_\tau$ functions breaks the
conformal symmetry and therefore one cannot fix three points by
symmetry. We find the same conclusion for our proposed Lagrangian.
It would be of interest to find a suitable conformally symmetric
Lagrangian that directly leads to the amplitudes \eqref{SuperAmps}.

As prospects for future work it is worth mentioning that the results obtained here can be generalized in various directions. The work done here was for $\mathbb{Q}_p$, but in principle one can
apply it to unramified extensions of the $p$-adic field
$\mathbb{Q}_{p^n}$. In the spirit of $p$-adic AdS/CFT this would
mean having multiple worldsheet coordinates. We have restricted the
value of $p$, yet the case $p=2$ remains to be explored.
$\mathbb{Q}_{2}$ admits antisymmetric sign functions, but they
behave very differently from their odd primed partners. Finding
more vector amplitudes like the ones proposed in \cite{MarshakovNew} can also be useful to understand better the
theory.

Another future direction is including a
$B$-field as in \cite{GhoshalBfield} in the context of superstrings. Recently it has been explored the idea of the $p$-adic bosonic
string as a $p2$-brane defined on the Bruhat-Tits tree
\cite{Huang:2021bsg}. Further generalizations of this idea require
the extension to the supersymmetric case. It would be very
interesting to pursue the consideration of the results presented in
our work to this notion of $p$-branes. Recently, the work \cite{Stoica:2021ncy}
found a relation between the bosonic 4- and 5- point amplitudes. It would be 
interesting to know if this relation can be extended to the superstring case.

Finally, very recently a rigorous study of the $p$-adic bosonic open string amplitudes has been 
carried out in \cite{Fuquen-Tibata:2021vnm}. In this reference it was shown that in the Euclidean case
the limit of a regularized Feynman integral is well defined and the standard non-Archimedean
Koba-Nielsen amplitudes are obtained only as the lower term of a series. In this series each term is well defined 
but the convergence of the series is still an open problem. It would be interesting to implement the procedure followed in 
\cite{Fuquen-Tibata:2021vnm} for the case of the superstring action considered in the present paper and look for a physical interpretation of those terms in the superstring context. 

 \vspace{1cm}
\centerline{\bf Acknowledgments} \vspace{.5cm} It is a pleasure to thank A. G\"uijosa and W. Z\'uñiga-Galindo by useful comments. E.Y.  L\'opez was
supported by the CONACyT graduate fellowship number 729722. 


\appendix

\section{Lightning review of $p$-adic numbers}\label{p-AdicStuff}
The rational numbers $\mathbb{Q}$ are topologically incomplete. Sequences of rational numbers exist that do not converge to a rational number (think of subsequently adding all of the digits of $\pi$ to $3.14$). One `fills in the gaps'' by adding such limits, this process requires the notion of convergence, that requires a norm. However not all norms are created equal, there are two types, the absolute value $|\cdot |$, and $p$-adic norms, denoted by $|\cdot |_p$. Throughout this entire paper $p$ stands for a prime number (2, 3, 5,\dots). For more details see, for instance, \cite{Vladimirov:1994wi}.

The field of $p$-adic numbers (denoted $\mathbb{Q}_p$) is defined as the completion of the rational numbers with respect to the $p$-adic norm. Consider a prime number $p$, and a rational number $r=\frac{a}{b}p^n$ with $a$ and $b$ coprime to $p$, and $n\in \mathbb{Z}$. The $p$-adic norm is defined as
\[|r|_p:= \begin{cases} p^{-n}, &r\neq 0\\ 0, &r=0 \end{cases}. \]
Notice that we have an infinite amount of $p$-adic norms, one per prime number. A $p$-adic number $x\neq0$ has a unique expansion 
\be x=p^{-v(x)}\sum_{m=0}^{\infty}x_m p^{m},\label{pSeries}\ee
with $x_m\in \{0,1,\dots,p-1\}; \ x_0\neq0$, $v(x)\in \mathbb{Z}$ is the valuation or order of $x$, and now $|x|_p=p^{v(x)}$. 
The unit ball is denoted by $\mathbb{Z}_p:=\{x\in \qp; \ |x|_p\leq 1\}$. This implies that $p^{k}\mathbb{Z}_p=\{x\in \qp; \ |x|_p\leq p^{-k}\}$, are the balls of radius $p^{-k}$ centered at $0$. The unit circle is $\mathbb{Z}_p^{\times}:=\{x\in \qp; \ |x|_p= 1\}$, that implies $p^k\mathbb{Z}_p^{\times}=\{x\in \qp; \ |x|_p= p^{-k}\}$.
We extend the $p-$adic norm to $
\mathbb{Q}
_{p}^{n}$ by taking
\[
||x||_{p}:=\max_{1\leq i\leq n}|x_{i}|_{p},\text{ for }x=(x_{1},\dots
,x_{n})\in
\mathbb{Q}
_{p}^{n}.
\]
We define $v(x)=\min_{1\leq i\leq n}\{v(x_{i})\}$, \ then \ $||x||_{p}%
=p^{-v(x)}$.\ The metric space $\left(
\mathbb{Q}
_{p}^{n},||\cdot||_{p}\right)  $ is a separable complete ultrametric space.
$p$-Adic balls in multiple dimensions are the product of one-dimensional balls, $\mathbb{Z}_p^n=\mathbb{Z}_p\times\mathbb{Z}_p\times\dots\times\mathbb{Z}_p$. However this doesn't happen for circles, $(\mathbb{Z}_p^2)^\times \neq \mathbb{Z}_p^\times\times \mathbb{Z}_p^\times$. From the definition one can see that in fact $(\mathbb{Z}_p^2)^\times = p\mathbb{Z}_p\times \mathbb{Z}_p^\times\sqcup\mathbb{Z}_p^\times\times p\mathbb{Z}_p \sqcup \mathbb{Z}_p^\times\times \mathbb{Z}_p^\times$, where $\sqcup$ is the union of disjoint sets.

\subsection{Integration}
As a locally compact topological group, $(\mathbb{Q}_{p},+)$ has a Haar measure $dx$, which is invariant under translations, i.e. $d(x+a)=dx$. If we normalize this measure by the condition $\int
_{\mathbb{Z}_{p}}dx=1$, then $dx$ is unique. Under scaling the measure behaves as $d(ax)=|a|_pdx$. The same properties hold for $(\mathbb{Q}_{p}^{n},+)$, where the Jacobian is used for the changes of variable.\\
 As a simple example we do the following integrations
\be\int_{p^k\mathbb{Z}_p}dx=\int_{\mathbb{Z}_p}d(p^k x)=p^{-k}\int_{\mathbb{Z}_p}dx=p^{-k};\label{Int1App}\ee
\[\int_{p^k\mathbb{Z}_p^\times}dx=\int_{p^k\mathbb{Z}_p}dx-\int_{p^{k+1}\mathbb{Z}_p}dx=p^{-k}(1-p^{-1}).\]
The discrete topology of $\mathbb{Q}_p$ implies $\displaystyle p^k\mathbb{Z}_p=\bigsqcup_{m=k}^{\infty}p^m\mathbb{Z}_p^\times$. From this we can see that
$$
\int_{p^k\mathbb{Z}_p}|x|_p^a dx=\sum_{m=k}^{\infty}|x|_p^{a}\int_{p^m\mathbb{Z}_p^\times}dx
$$
$$
=\sum_{m=k}^{\infty}p^{-m(a+1)}(1-p^{-1})
$$
$$
=\frac{p^{-k(a+1)}}{1-p^{-(a+1)}}(1-p^{-1}),
$$
if we set $a=0$ the result is the same as in \eqref{Int1App}.

\subsection{The sign function}
We first define the following function known as the Legendre symbol
\[
\bigg(\frac{a}{p}\bigg)=\left\{
\begin{array}
[c]{lll}%
1 & \text{if} & x^{2}\equiv a \ {\rm mod} \ p\ \text{has\ a\ solution}\\
&  & \\
-1 & \text{if} & \text{otherwise,}%
\end{array}
\right.
\]
where $a$ is an integer. This may be thought of as the sign function for the finite field of $p$ elements $\mathbb{F}_p$. Now let $\left[  \mathbb{Q}_{p}%
^{\times}\right]  ^{2}$ be the multiplicative subgroup of squares in
$\mathbb{Q}_{p}^{\times}$, i.e.
\[
\left[  \mathbb{Q}_{p}^{\times}\right]  ^{2}=\{a\in\mathbb{Q}_{p}%
;a=b^{2}\text{ for some }b\in\mathbb{Q}_{p}^{\times}\}.
\]
For $p\neq2$, and $\epsilon\in\{1,\ldots,p-1\}$ satisfying $(\frac
{\epsilon}{p})=-1$, we have
\[
\mathbb{Q}_{p}^{\times}/\left[  \mathbb{Q}_{p}^{\times}\right]  ^{2}%
=\{1,\epsilon,p,\epsilon p\}.
\]
Then any nonzero $p$-adic number can be written uniquely as follows
\[
x=\tau a^{2},\text{ with }a\in\mathbb{Q}_{p}^{\times}\text{ and }\tau
\in\mathbb{Q}_{p}^{\times}/\left[  \mathbb{Q}_{p}^{\times}\right]  ^{2}.
\]
Take a fixed $\tau\in\{\epsilon,p,\epsilon p\}$, and $x\in\mathbb{Q}%
_{p}^{\times}$, the usual definition of the sign function is given by
\begin{equation}
\mathrm{sgn}_{\tau}(x):=%
\begin{cases}
1 & \text{if}\ x=a^{2}-\tau b^{2}\ \text{for}\ a,b\in\mathbb{Q}_{p}\\
-1 & \text{otherwise.}%
\end{cases}
\label{signdef}%
\end{equation}
All the possible $p$-adic sign functions are better summarized in the following table (see \cite{Gubser:2018cha}):%
\begin{equation}%
\begin{tabular}
[c]{|l|l|}\hline
$p\equiv1$ $\operatorname{mod}$ $4$ & $p\equiv3$ $\operatorname{mod}$
$4$\\\hline
$\mathrm{sgn}_{\epsilon}(x)=\left(  -1\right)  ^{v(x)}$ & $\mathrm{sgn}%
_{\epsilon}(x)=\left(  -1\right)  ^{v(x)}$\\\hline
$\mathrm{sgn}_{p}(x)=\left(  \frac{x_{0}}{p}\right)  $ & $\mathrm{sgn}%
_{p}(x)=\left(  -1\right)  ^{v(x)}\left(  \frac{x_{0}}{p}\right)  $\\\hline
$\mathrm{sgn}_{\epsilon p}(x)=\left(  -1\right)  ^{v(x)}\left(
\frac{x_{0}}{p}\right)  $ & $\mathrm{sgn}_{\epsilon p}(x)=\left(
\frac{x_{0}}{p}\right)  .$\\\hline
\end{tabular}
\ \label{Table}%
\end{equation}

\subsection{Fourier Transform}\label{AppFourier}
Fourier analysis is very similar to the usual Archimedean case. The $p$-adic Fourier transform of a locally constant function $\phi(x)$ is defined as 
\be \tilde{\phi}(\omega)=\pint \chi(\omega x) \phi(x)dx, \ee
where $\chi(x)=e^{2\pi i \{x\}_p}$, with $\{x\}_p$ being the fractional part of $x$, i.e. the terms with negative powers of $p$ in \eqref{pSeries}. One can show the following
\[\int_{p^k \mathbb{Z}_p}\chi(\omega x)dx=\begin{cases}p^{-k}, &|\omega|_p\leq p^k \\ 0, &|\omega|_p\geq p^{k+1} \end{cases},\]
which is used to prove that
\[\delta(x)=\pint \chi(\omega x)d\omega. \]
With this we can obtain the inverse transformation
\be\phi(x)=\pint \chi^*(x\omega) \tilde{\phi}(\omega)d\omega.\ee
\section{Vertex operators} \label{AppVertex}

We briefly review the process for the basic tree amplitudes in the
Archimedean superstrings done in  Ref. \cite{GSWSuperstrings}. Consider
the vertex operator \be V(k;X,\psi)=k\cdot \psi :e^{ik\cdot X}
=\int d\theta e^{ik\cdot X + \theta k\cdot \psi}, \label{VertexOps}
\ee where $\theta$ is an auxiliary Grassmann variable. It is important to note that the second equality above is a consequence of the Grassmann variables properties.\\
Now we use the two-point functions \cite{GSWSuperstrings} \ba
\left\langle X^\mu (y_i)X^\nu (y_j)\right\rangle &
= &-\eta^{\mu\nu}\log (y_i-y_j),\\
\left\langle \frac{\psi^\mu (y_i)}{\sqrt{y_i}}\frac{\psi^\nu
(y_j)}{\sqrt{y_j}}\right\rangle & = &{\displaystyle
\frac{\eta^{\mu\nu}}{y_i-y_j}}, \ea and we get that \ba \left\langle
\frac{V(k_i;y_i)}{\sqrt{y_i}}\frac{V(k_j;y_j)}{\sqrt{y_j}}\right\rangle & = \int
d\theta_i d\theta_j
e^{k_i\cdot k_j \left( \log(y_i-y_j)-\frac{\theta_i\theta_j}{y_i-y_j} \right)} \nonumber \\
&= \int d\theta_i d\theta_j (y_i-y_j - \theta_i\theta_j)^{k_i\cdot
k_j}. \ea Now for multiple vertex operators we have \be
\prod_{l=1}^{N} V(k_l;y_l) =\int  d\theta_1 \cdots d\theta_N
\exp\left\{\sum_{l=1}^N
\boldsymbol{k}_l\cdot\left(i\boldsymbol{X}(y_l)+\theta_l\boldsymbol{\psi}(y_l)\right)\right\}.
\ee Then \bs \left\langle\prod_{l=1}^{N}
\frac{V(k_l;y_l)}{\sqrt{y_l}}\right\rangle &=\int  d\theta_1 \cdots
d\theta_N\\
&\times \exp\left\{\sum_{l,m=1}^N k_{l,\mu}k_{m,\nu}\left(-\big\langle X^\mu(y_l)X^\nu(y_m)\big\rangle-\theta_l\theta_m \bigg\langle\frac{\psi^\mu(y_l)}{\sqrt{y_l}}\frac{\psi^\nu(y_m)}{\sqrt{y_m}}\bigg\rangle\right)\right\}  \\
&=\int  d\theta_1 \cdots d\theta_N
\exp\left\{\sum_{l,m=1}^N \boldsymbol{k}_{l}\cdot \boldsymbol{k}_{m}\left(\log(y_l-y_m)-\frac{\theta_l\theta_m}{y_l-y_m}\right)\right\}\\
&=\int  d\theta_1 \cdots d\theta_N \prod_{l<m}(y_l-y_m -
\theta_l\theta_m)^{\boldsymbol{k}_{l}\cdot \boldsymbol{k}_{m}}.
\end{split}\label{ArchAmp}
\ee
This derivation demanded
only Fubini's theorem and changes of variables, both are well defined
over the $p$-adics. Thus in the non-Archimedean setting, we can follow this same path, the main difference would be the two point function, that is described in the main text.\\
As a side note, the last equality of \eqref{ArchAmp} used the following identity for Grassmann variables $\theta_i$
 \ba \log(y_i-y_j)
+\frac{\theta_i\theta_j}{y_i-y_j}=\log (y_i-y_j)
+\log\bigg(1+\frac{\theta_i\theta_j}{y_i-y_j}\bigg)   \nonumber \\
=\log(y_i-y_j +\theta_i\theta_j). \ea This is actually quite
general, in fact one can easily check that for constants $A,B$ and $s$, the following is true \ba
A\log|y_i-y_j|+B \frac{\theta_i\theta_j}{|y_i-y_j|^s}=\frac{A}{s}\left( \log|y_i-y_j|^s+\frac{Bs}{A}\frac{\theta_i\theta_j}{|y_i-y_j|^s}\right) \nonumber \\
=\frac{A}{s}\log\left( |y_i-y_j|^s + \frac{Bs}{A}\theta_i\theta_j\right).
\label{LogSum} \ea
This more general identity is used in the non-Archimedean case.
\section{Functional derivatives}\label{AppDerivative}

In this appendix we define in more detail a functional derivative for $p$-adic fermion fields. It is done in a very
similar way to the usual bosonic variables. We also use it to obtain the fermion propagator as a two point function. Even though we are using Grassmann variables, commutativeness issues do not arise because we use only pairs of Grassmann variables.\\
We define the functional derivative
for the Grassmann field $K$ \be \frac{\delta Z[K]}{\delta K^\mu (y)}=\int
d\theta \lim_{\varepsilon\to 0}\frac{Z[K+\varepsilon\theta
\delta^\cdot_\mu \delta(\cdot - y)]-Z[K]}{\varepsilon}, \label{FuncDer}
\ee where $\theta$ is a Grassmann variable. The dots indicate a missing argument in the deltas. Consider the following partition function with a propagator
$G_{\mu\nu}(x)$ that is antisymmetric (it satisfies $G_{\mu\nu}(-x)=-G_{\mu\nu}(x)$)
\be Z[K]=\exp\left\{
\frac{1}{2}\int_{\qp^2}K^\mu(x) G_{\mu\nu}(x-y)K^\nu(y)dxdy \right\}.
\ee Now let's first see that \small
$$
 Z[K+\varepsilon\theta \delta^\cdot_\alpha
\delta(\cdot - z)] =\exp\left\{
\frac{1}{2}\int_{\qp^2}(K^\mu(x)+\varepsilon\theta\delta^\mu_\alpha
\delta(x-z)
)G_{\mu\nu}(x-y)(K^\nu(y)+\varepsilon\theta\delta^\nu_\alpha
\delta(y-z))dxdy \right\}
$$
\normalsize
$$
=\exp\left\{  \frac{1}{2}\int_{\qp^2}K^\mu(x)
G_{\mu\nu}(x-y)K^\nu(y)dxdy \right\}
$$
$$
\times \exp\left\{
\frac{1}{2}\varepsilon\int_{\qp^2}\left[K^\mu(x)G_{\mu\nu}(x-y)\theta\delta^\nu_\alpha
\delta(y-z)+\theta\delta^\mu_\alpha
\delta(x-z)G_{\mu\nu}(x-y)K^\nu(y)\right]dxdy  \right\}
$$
$$
=Z[K]\exp\left\{  \frac{1}{2}\varepsilon\pint
G_{\mu\alpha}(x-z)\left[K^\mu(x)\theta-\theta K^\mu(x)\right]dx
\right\}
$$
$$
=Z[K]\exp\left\{  \varepsilon\theta\pint G_{\mu\alpha}(z-x)K^\mu(x)dx    \right\}\nn \\
$$
\be =Z[K]\left[1+\varepsilon\theta\pint G_{\mu\alpha}(z-x)K^\mu(x)dx +
\mathcal{O}(\varepsilon^2) \right]. \ee Thus, we can now obtain the
functional derivative
$$
\frac{\delta Z[K]}{\delta K^\mu (y)}=\int d\theta \lim_{\varepsilon\to
0} \left[\theta\pint G_{\mu\alpha}(z-x)K^\mu(x)dx Z[K]+
\mathcal{O}(\varepsilon) \right]
$$
\be =\pint G_{\mu\alpha}(y-x)K^\mu(x)dx Z[K].\ee Our functional
derivative \eqref{FuncDer} follows the Leibniz rule, one can easily
check this. Now we can obtain the two point function
$$
\frac{\delta^2 Z[K]}{\delta K^\mu (y_1)\delta K^\nu
(y_2)}\Bigg\rvert_{K=0} =\bigg[ G_{\mu\nu} (y_1-y_2)+\pint
G_{\mu\alpha}(y_1-x)K^\alpha(x)dx
$$
\be \times \pint G_{\nu\beta}(y_2-x)K^\beta(x)dx \bigg]
Z[K]\Bigg\rvert_{K=0} =G_{\mu\nu} (y_1-y_2). \ee One also can
easily check that \be \frac{\delta}{\delta K^\mu
(y)}\exp\left\{\pint K_\nu(x) \psi^\nu (x)
dx\right\}\Bigg\rvert_{K=0}=\psi_\mu (x). \ee 
Looking at the fermionic part in \eqref{GenFunction} coming from the action $I_F[\psi]$. We see that indeed \be
\langle\psi^{\mu}(x)\psi^{\nu}(y)\rangle=\frac{\a}{\st(-1)p}\G_\fer^{\mu\nu}(x-y). \ee

\newpage


\end{document}